\documentstyle[preprint,tighten,aps,epsf]{revtex}
\def\lsim{\raise0.3ex\hbox{$\;<$\kern-0.75em\raise-1.1ex\hbox{$\sim\;$}}}
\def\gsim{\raise0.3ex\hbox{$\;>$\kern-0.75em\raise-1.1ex\hbox{$\sim\;$}}}
\begin{document}
\preprint{SISSA/EP/140/98~~~IC/98/228}
\draft
\title{Fully Supersymmetric CP Violation in K and B Systems}
\author{D. A. Demir}
\address{The Abdus Salam International Center for Theoretical Physics, I-34100,
Trieste, Italy }
\author{A. Masiero and O. Vives}
\address{SISSA -- ISAS, Via Beirut 4, I-34013, Trieste, Italy and \\ INFN,
Sezione di Trieste, Trieste, Italy.}
\maketitle

\begin{abstract}
We analyze CP violation in supersymmetric extensions of the Standard Model 
with heavy scalar fermions of the first two generations. Neglecting 
intergenerational mixing in the sfemion mass matrices and thus considering only 
chargino, charged Higgs and W--boson diagrams we show that it is possible to fully 
account for CP violation in the kaon system even in the absence of the 
standard CKM phase. This opens new possibilities for large supersymmetric 
contributions to CP violation in the B system.   

\end{abstract}
Beginning with its experimental discovery in K--meson decays, about 
three decades ago, the origin of CP violation has been one of the most 
intriguing questions in particle phenomenology. Notably, the subsequent 
experiments in the search for electric dipole moments (EDM) of the neutron and 
electron have observed no sign of new CP--violating effects despite their 
considerably high precision. 
However, this situation is expected to change in the near future with 
the  advent of the new B factories. Indeed, experimental studies of the 
neutral B--meson systems can provide a window to new physics beyond the 
Standard Model (SM) much earlier than the direct collider searches at the LHC. 

In the standard electroweak theory the CP--violating phenomena find
their explanation uniquely in the phase $\delta_{CKM}$ of the
Cabibbo--Kobayashi--Maskawa mixing matrix (CKM). However, in the minimal 
supersymmetric extension of the SM (MSSM) there are additional phases 
which can cause deviations from the predictions of the SM. Indeed, 
after all possible rephasings of the parameters and fields there 
remain two new physical phases in the  MSSM Lagrangian that can be chosen
to be  
the phases of the Higgsino Dirac mass parameter 
($\varphi_{\mu}=\mbox{Arg}[\mu]$) and  the trilinear sfermion coupling 
($\tilde{f}$) to the Higgs, ($\varphi_{A_{f}}=\mbox{Arg}[A_{f}]$) 
\cite{2phases}.
In the absence of 
the strict universality, as is generally the case at low energies, each 
sfermion species has a distinct phase $\varphi_{A_{f}}$. In the presence of 
such CP--violating phases a natural question to be raised would be 
``Is it possible to account for the observed CP--violation only by 
supersymmetric effects ?''. 

The traditional answer to this question has always been negative because of
the fact that the electric dipole moments of the electron and 
neutron constrain $\varphi_{A_f,\mu}$ to be at most ${\cal{O}}(10^{-2})$. 
However, recent studies have revealed new pathways to small enough EDM's 
while allowing SUSY phases ${\cal{O}}(1)$. Methods of suppressing the EDM's 
consist of cancellation of various SUSY contributions among themselves 
\cite{cancel}, non-universality of the soft breaking parameters at the 
unification scale \cite{abel}, and approximately degenerate heavy sfermions 
for the first two generations \cite{heavy}. 
In this Letter we shall follow the last alternative and assume a general low 
energy SUSY model with heavy and almost degenerate sfermions for the first 
two generations. Now the question at 
the end of the last paragraph takes a more specific form: ``In such a scheme, 
can $\varphi_{A_f}$ and $\varphi_{\mu}$ account for the experimental 
observation of CP violation with vanishing $\delta_{CKM}$?''.

To investigate the answer to this question we follow a simplifying assumption,
that is, we neglect all intergenerational mixings in the sfermion mass 
matrices. Then, neutralino and gluino vertices are approximately flavor 
diagonal and these particles do not contribute to the flavor changing neutral 
current processes. Under this assumption, all flavor mixing effects 
originate from the elements of CKM matrix, as in the SM. In this scenario, 
$\Delta F= 2$ transitions proceed only through box diagrams exchanging 
$(W^{\pm})$--quarks, charged Higgs boson ($H^{\pm}$)--quarks, and 
chargino ($\chi^{\pm}$)--squarks. The first contribution is the usual SM 
while the other two are of SUSY origin. These three types of 
contributions are always present independently of the existence/absence of 
intergenerational mixings, and thus the results we present in this Letter 
can be regarded as conservative limits on SUSY effects. 

Under the conditions mentioned above it turns out that 
\begin{enumerate}
\item Relevant new SUSY contributions to CP violation observables 
are only possible when the masses of the light chargino and stop are 
close to the $Z$ mass ($M_{\chi_{1}}$, $M_{\tilde{t}_{1}} \lsim 150 GeV$), 
and $\tan\beta \gsim 30$.
\item In such case, the answer to our question is definitely positive,
$\varepsilon_K$ and the corresponding observable in the B system, 
$\varepsilon_B$, get large contributions from SUSY that can even saturate 
the measured value for $\varepsilon_K$.                    
\end{enumerate}
 
The $K^0$--$\bar{K}^0$ and $B^0$--$\bar{B}^0$ mixings are conveniently 
described by the  corresponding $\Delta F =2$ effective Hamiltonian 
\begin{eqnarray}
\label{DF=2}
{\cal{H}}_{eff}^{\Delta F=2}=-\frac{G_{F}^{2} M_{W}^{2}}{(2 \pi)^{2}}
(V_{td}^{*} V_{tq})^{2}( C_{1}(\mu) Q_{1}(\mu)
+C_{2}(\mu) Q_{2}(\mu) +C_3(\mu) Q_3(\mu))
\end{eqnarray}
where $Q_{1}=\bar{d}^{\alpha}_{L}\gamma^{\mu}q^{\alpha}_{L}\cdot 
\bar{d}^{\beta}_{L}\gamma_{\mu}q^{\beta}_{L}$, $Q_{2}=\bar{d}^{\alpha}_{L}
q^{\alpha}_{R}\cdot \bar{d}^{\beta}_{L}q^{\beta}_{R}$, $Q_{3}=
\bar{d}^{\alpha}_{L}q^{\beta}_{R}\cdot \bar{d}^{\beta}_{L}q^{\alpha}_{R}$ 
are the effective four--fermion operators with $q=s , b$ for the $K$ and 
$B$--systems respectively, $\alpha, \beta$ are color indices and $C_{1,2,3}$ 
the corresponding Wilson coefficients. 
These are evaluated at the corresponding meson mass, $\mu$, with the initial 
conditions specified at $\mu_{0}\sim M_{\tilde{t}}$. Without the inclusion of 
QCD corrections, ${C_{2}}(\mu_{0})$ receives no contribution from $W^\pm$, 
charged Higgs and chargino boxes, and the other two coefficients can be 
decomposed according to the particles in the loop as 
${C_{1}}(\mu_{0})={C_{1}^{W}}(\mu_{0})+{C_{1}^{H}}(\mu_{0})+
{C_{1}^{\chi}}(\mu_{0})$, ${C_{3}}(\mu_{0})={C_{3}^{H}}(\mu_{0})+
{C_{3}^{\chi}}(\mu_{0})$. The contributions of 
$W$ boson, ${C_{1}^{W}}(\mu_{0})$, and charged Higgs, 
${C_{1,3}^{H}}(\mu_{0})$, are real. 

The chargino boxes are the only genuine SUSY contribution to the mixing.
The explicit expressions for their contribution to the Wilson coefficients
are,

\begin{eqnarray}
\label{chWC}
C_1^\chi (\mu_{0}) = \frac{1}{4} \sum_{i,j=1}^{2} \sum_{k, l=1}^{2} G^{(3,k)i} G^{(3,k)j*} G^{(3,l)i*} 
 G^{(3,l)j}\  Y_1(z_k, z_l, s_i, s_j) \\
C_3^\chi (\mu_{0}) = \sum_{i,j=1}^{2} \sum_{k, l=1}^{2} H^{(3,k)i} G^{(3,k)j*} G^{(3,l)i*}  H^{(3,l)j}\  Y_2(z_k, z_l, s_i, s_j) \nonumber 
\end{eqnarray}

where $z_k = M_{\tilde{t}_k}^2/M_W^2$, $s_i =M_{\tilde{\chi}_i}^2/M_W^2$, and 
the stop--chargino--quark coupling matrices $G^{(3,k)i}$ and $H^{(3,k)i}$ are 
combinations of the stop ($S_t$) and chargino ($C_{R,L}$) mixing matrices 
\begin{equation}
\label{couplings}
G^{(3,k)i}= \sqrt{2} C^*_{R 1 i}S_{t k 1} - \frac{C^*_{R 2 i}S_{t k 2}}
{\sin \beta} \frac{m_t}{M_W} , \hspace{1cm} H^{(3,k)i} = \frac{C^*_{L 2 i} S_{t k 1}}
{\cos \beta} \frac{m_q}{M_W}\; .
\end{equation}
The complete expressions for the loop functions $Y_{1,2}(a,b,c,d)$ as well as 
$W^{\pm}$ and charged Higgs contributions can be found in \cite{branco}. 
The matrices $G^{(3,k)i}$ and $H^{(3,k)i}$ represent the coupling of chargino 
and stop to left and right--handed down quarks respectively. 
As can be seen from Eq.(\ref{chWC}), unlike $C_1^\chi (\mu_{0})$, 
$C_3^\chi (\mu_{0})$ depends on both $G^{(3,k)i}$ and $H^{(3,k)i}$, and thus,
in general, it is complex. On the other hand, unless one admits large enough
$\tan\beta$ values, $|H^{(3,k)i}|$ becomes much smaller than $|G^{(3,k)i}|$ 
since the former is suppressed by the ratio of $m_{q=b,s}$ to $M_{W}$.

Solution of RGE's at the corresponding meson mass scale yield \cite{RGE}
\begin{equation}
\label{RGE}
C_1(\mu)= 0.790\; C_1(\mu_{0}) \;\;,\;\;C_2(\mu)= - 0.056\; C_3(\mu_{0})\;\;,
\;\; C_3(\mu)= 2.930\; C_3(\mu_{0})
\end{equation}  
where the enhancement of the $C_3(\mu)$ Wilson coefficient is specially 
interesting.  

With the complete expression of the $\Delta F=2$ effective Hamiltonian, it 
is now a straightforward issue to analyze CP--violation observables. 
For both K and B--meson systems we follow the usual definition
of the $\varepsilon$ parameter factoring out global phases,
\begin{eqnarray}
\label{ek}
\varepsilon_{{\cal{M}}}=\frac{1}{\sqrt{2}}\frac{{\cal{I}}m\langle {\cal{M}}^0 | {\cal
H}_{eff}^{\Delta
F=2} |\bar{{\cal{M}}}^0 \rangle}{\Delta M_{\cal{M}}}
\end{eqnarray}
where ${\cal{M}}^{0}=K^0, B^0$, and correspondingly, $q=s, b$ in the effective Hamiltonian. The off--diagonal matrix element of the effective Hamiltonian 
in the neutral meson mass matrix is
\begin{eqnarray}
\label{m12}
\langle {\cal{M}}^0 | {\cal H}_{eff}^{\Delta F=2} | \bar{{\cal{M}}}^0 \rangle
= -\frac{G_{F}^{2} M_{W}^{2}}{(2 \pi)^{2}} (V_{td}^{*} V_{tq})^{2}\
F_{{\cal{M}}}^{2} M_{{\cal{M}}} \{\frac{1}{3} C_{1}(\mu) B_1(\mu)\nonumber\\ 
+ \left(\frac{M_{{\cal{M}}}}{m_q(\mu) + m_d(\mu)} \right)^2 (
- \frac{5}{24} C_{2}(\mu) B_2(\mu) + \frac{1}{24} C_{3}(\mu) B_3(\mu))\}
\end{eqnarray} 
In this expression $M_{{\cal{M}}}$ and $F_{{\cal{M}}}$ denote
the mass and decay constant of the neutral meson ${\cal{M}}^{0}$, 
$m_{q=s,b}(\mu)$, and $m_{d}(\mu)$ are the quark masses in the corresponding 
renormalization scheme at the scale $\mu$. The $B$--parameters at $\mu=2 GeV$
are: $B_1(\mu)=0.60$, $B_2(\mu)=0.66$ and $B_3(\mu)=1.05$ \cite{RGE}.
 
Using Eq.(\ref{ek}) we can easily calculate the amount of indirect 
CP--violation in the K and B--systems. As Eqs.(\ref{ek}) and (\ref{m12}) 
suggest $\varepsilon_{{\cal{M}}}$ gets contributions from two different 
sources: the usual CKM phase $\delta_{CKM}$ coming through 
$(V_{td}^{*} V_{tq})^{2}$, and the complex Wilson coefficients, $C_{2,3}(\mu)$.
While the former is the usual source of CP--violating phenomena in the SM 
the latter is completely a SUSY effect. In the following, we will
{\it always assume that there is no physical phase in the CKM mixing matrix,
and CP violation is purely of Supersymmetric origin}.
  
In first place we will analyze the $K$--system where we have experimental 
confirmation of CP--violation. At first sight, one would completely neglect
any SUSY contribution to $\varepsilon_{K}$ because of the huge suppression
factor associated with the $s$--quark Yukawa coupling in $H^{(3,k) i}$. 
However, this is not necessarily true for large $\tan \beta$ (with $\tan 
\beta= 60$ we get $(m_s/(M_W \cos \beta))^2 ={\cal O}(10^{-2})$). 
Besides this, we have additional enhancements from the hadronic matrix 
elements, $\langle \bar{K}^0 | Q_3 | K^0 \rangle \approx 3 \cdot 
\langle \bar{K}^0 | Q_1 | K^0 \rangle$ and the RGE evolution, Eq.(\ref{RGE}).
At this point, it is clear that it is not correct to 
neglect the $C_3$ contribution to the imaginary part of $M_{1 2}$ 
for large $\tan \beta$. 
Nevertheless sizeable effects can be expected only for light enough stop and 
chargino. In Fig. 1 we show $\varepsilon_K$ as a function
of $\varphi_\mu$ and $\varphi_{A_t}$ for $\tilde{M}_2=\mu=125\ GeV$
, $M_{\tilde{t}_L}=M_{\tilde{t}_R}=150\ GeV$, $A_t=250\ GeV$ and 
$\tan \beta = 60$ which give on average $M_{\tilde{t}_1} = 83\ GeV$ and 
$M_{\tilde{\chi}_1} = 80\ GeV$. As we can see in this figure, it is still 
possible to fully saturate the measured value of $\varepsilon_{K}$ relying
only on SUSY phases.

Indirect CP--violation in the B--system can be analyzed in the same way from
the $\Delta B=2$ effective Hamiltonian, Eq.(\ref{DF=2}). The effects in the 
corresponding observable $\varepsilon_{B}$ will increase roughly as 
$(m_b/m_s)^2$ with respect to $\varepsilon_{K}$. In Fig. 2 we show 
$\varepsilon_{B}$ as a function of 
$\varphi_\mu$ and $\varphi_{A_t}$ for the same region of SUSY parameter 
space that we used in the $\varepsilon_K$ analysis. Notice that for the set 
of values for which $\varepsilon_K$ is fully saturated we obtain large 
effects for CP violation in the B system. In this system, the 
mixing--induced CP phase, $\theta_M$, measurable in $B^0$ CP asymmetries,
is related to $\varepsilon_B$ by $\theta_M=\arcsin\{2 \sqrt{2} 
\cdot \varepsilon_B \}$. For $\varepsilon_B = 0.2$, it reaches the value 
$\theta_M\approx 0.6$ which can be cleanly observed in the future B factories.
Even for smaller values of 
$\tan \beta$, for which $\varepsilon_K$ is not entirely reproduced by SUSY 
effects, we still have quite sizeable SUSY contributions to CP violation in B
mixing with appreciable deviations from the SM results \cite{WIP}.

In conclusion, we have shown in this letter that in a general SUSY extension 
of the SM with heavy scalar fermions of the first two generations it is 
possible to fully account for the observed CP violation in the kaon system.
Also in these SUSY models with phases ${\cal O}(1)$ we can expect large 
SUSY contributions to CP violation in the B system.

We thank S. Bertolini, I. Scimemi and E. Lunghi for enlightening discussions.
The work of A.M. was partially supported by the European TMR Project
``Beyond the Standard Model'' contract N. ERBFMRX CT96 0090; O.V. 
acknowledges financial support from a Marie Curie EC grant 
(TMR-ERBFMBI CT98 3087).

\section{ERRATUM}

In this letter, we analyzed indirect CP violation in a supersymmetric model 
with heavy sfermions of the first two generations and large susy phases.
We assumed that no flavor change occurs in the squark mass matrices.
The dependence of the new CP violating FCNC on the down quark Yukawa 
couplings implies that only the region of large $\tan \beta$ can give rise 
to observable effects.

However this region is strongly constrained by the $b \rightarrow s \gamma$
measurement. Specifically, after taking into account all the supersymmetric 
contributions, the example given in this letter with $\tan \beta= 60$ and 
large mixings in the chargino and sfermion matrices gives rise to a too large 
contribution to $b \rightarrow s \gamma$ excluded by the experimental 
measurement. Similarly, constrains from Electric Dipole Moments are again 
relevant for this large $\tan \beta$ region and must be considered 
\cite{2loop}.

Moreover when recalculating possible effects in $K^0$--$\bar{K}^0$ mixing 
we discovered a mistake in the program used to generate $\varepsilon_K$
that implied a further reduction of the presented values.

The presence of these strong constrains lead us to reconsider the conclusions
reached in this letter. A complete analysis will be presented in a subsequent 
work \cite{WIP}.

\clearpage

\begin{figure}
\begin{center}
\epsfxsize = 15cm
\epsffile{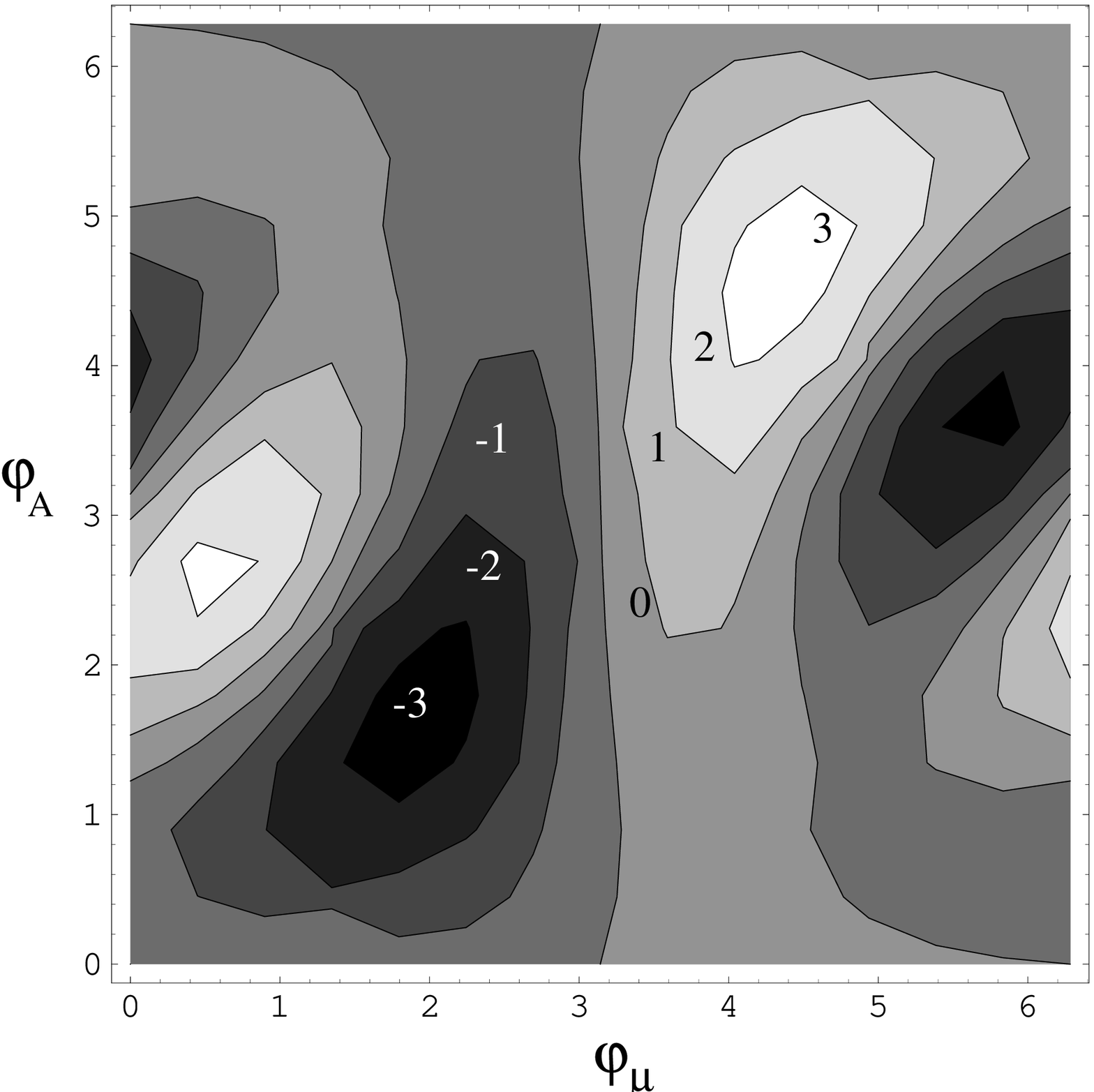}
\leavevmode
\end{center}
\caption{Values of $10^{3}\cdot \varepsilon_K$ in the $\varphi_\mu$, 
$\varphi_{A_f}$ plane for the region of SUSY parameter space specified 
in the text}
\end{figure}

\clearpage

\begin{figure}
\begin{center}
\epsfxsize = 15cm
\epsffile{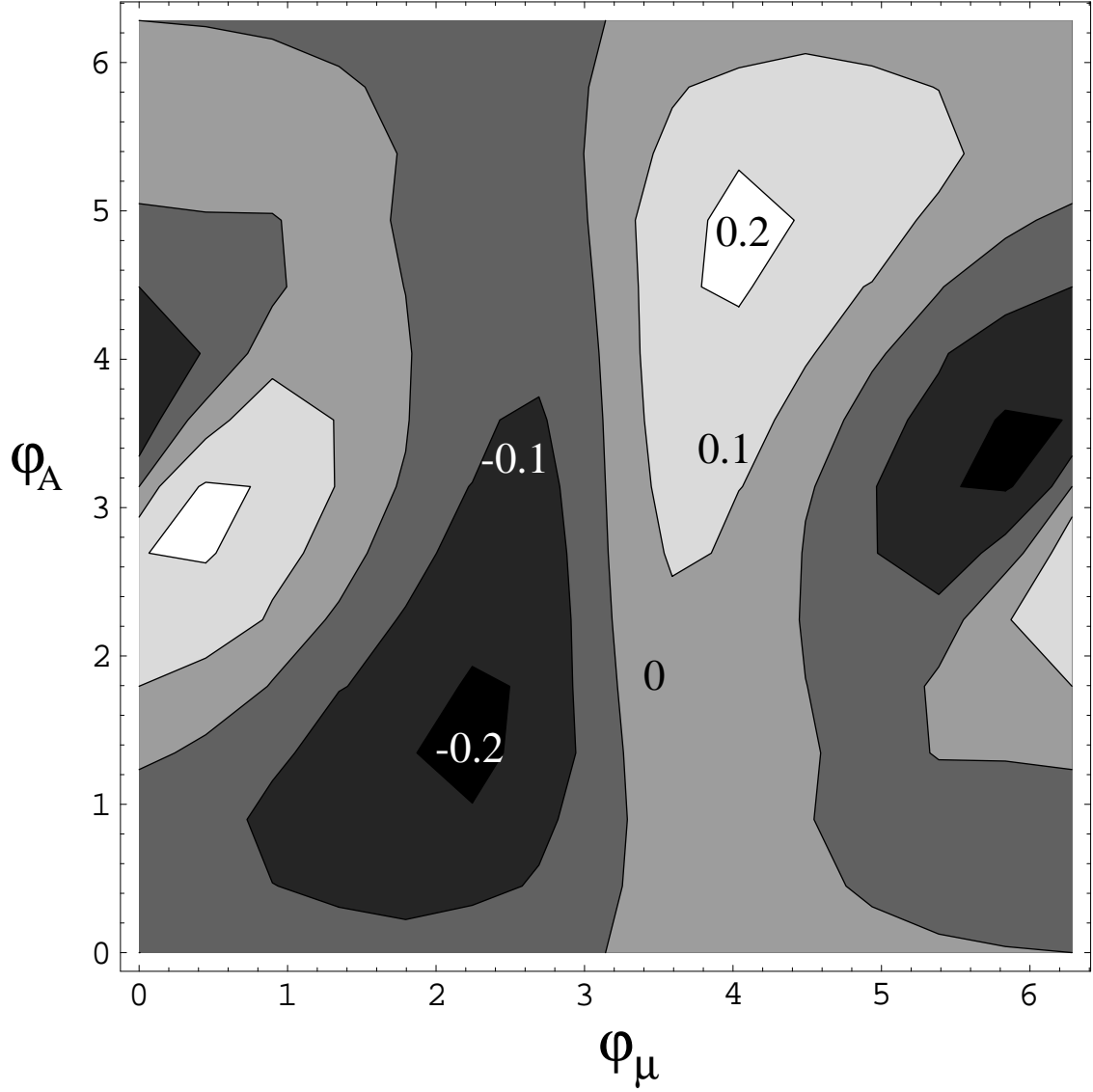}
\leavevmode
\end{center}
\caption{Values of $\varepsilon_B$ in the $\varphi_\mu$, 
$\varphi_{A_f}$ plane for the same region of SUSY parameter 
space as in Fig. 1}
\end{figure}
\end{document}